\documentclass[conference]{IEEEtran}
\IEEEoverridecommandlockouts

\usepackage{cite}
\usepackage{amsmath,amssymb,amsfonts}
\usepackage{algorithmic}
\usepackage{graphicx}
\usepackage{textcomp}
\usepackage{xcolor}    
\def\BibTeX{{\rm B\kern-.05em{\sc i\kern-.025em b}\kern-.08em
    T\kern-.1667em\lower.7ex\hbox{E}\kern-.125emX}}

\makeatletter
\def\ps@IEEEtitlepagestyle{%
  \def\@oddfoot{\mycopyrightnotice}%
  \def\@evenfoot{}%
}
\def\mycopyrightnotice{%
  {\footnotesize  %978-1-6654-8491-6/22/\$31.00 @2022 IEEE
  \hfill}% <--- Change here
  \gdef\mycopyrightnotice{}% just in case
}
\makeatother

\begin{document}

\title{Thermodynamic Model for the Oxygen Ion Mobility in Dense Argon Gas}

\author{\IEEEauthorblockN{%1\textsuperscript{st} 
 \underline{Armando Francesco Borghesani}}
\IEEEauthorblockA{\textit{CNISM Unit-Dept. of Physics \& Astronomy} \\
\textit{University of Padua}\\
Padua, Italy \\
armandofrancesco.borghesani@unipd.it}
\and
\IEEEauthorblockN{%2\textsuperscript{nd} 
Fr{\'e}d{\'e}ric Aitken}
\IEEEauthorblockA{\textit{G2ELab C.N.R.S. } \\
\textit{University Grenoble Alpes}\\
Grenoble, France \\
frederic.aitken@g2elab.grenoble-inp.fr}}

\maketitle

\begin{abstract}
We report experimental data of the mobility of O\(_2^-\) ions in argon gas as a function of the density in the temperature range between \(180\,\mbox{K}\le T\le 300\,\mbox{K}\). At the intermediate, though fairly large, densities of the experiment both the kinetic theory and the hydrodynamic theory fail at describing the experimental data. By contrast, the free volume model, originally developed to describe electron and ion mobility in superfluid helium, gives a satisfactory agreement with the  data as previously obtained for the O\(_2^-\) ion mobility in both dense neon and helium gases.
 \end{abstract}


\section{Introduction}
Fundamental research on the microscopic interactions of charges and neutral species as well as applications involving, for instance, low-temperature plasmas, require a detailed knowledge of how ions drift in a dense disordered medium.  Ion transport occurs in many instances such as chemical synthesis, high-energy particle detection, electrical discharges, atmospheric physics and more. In order to correctly design technical apparatuses it is very important to accurately model the transport properties of the ions in the  medium~\cite{Bruggeman2016b,lopez2005,mason2001}.

In the dilute gas regime, the classical kinetic theory yields an accurate prediction of the ion mobility \(\mu\) provided that the ion-atom scattering cross section is known so that the relevant collision integrals can be computed~\cite{viehland1975,viehland1975b,maitland,mason1988}. 
If the neutral-ion interaction can be modeled as a hard-sphere interaction of radius \(R_0\), the density-normalized mobility \(\mu N\)  of thermal ions, as are those in the case of the present experiment, is given by
\begin{equation} 
\mu N = \frac{3}{2\sigma}\left(\frac{\pi}{2m_r k_\mathrm{B}T}\right)^{1/2}
\label{eq:1}\end{equation} in which \(\sigma= 4\pi R_0^2\) is the scattering cross section, \(N\)\ and \(T\) are the gas density and temperature, \(k_\mathrm{B}\)\ is the Boltzmann constant, and \(m_r\) is the ion-neutral reduced mass. 

At the other end of the thermodynamic spectrum, there is the transport of ions in liquids. In this case, the mobility of thermal ions is described by the Stokes hydrodynamic formula
\begin{equation}
\mu N= \frac{eN}{6\pi \eta R}\label{eq:2}
\end{equation}
in which \(e\) is the ion charge and \(\eta\) is the liquid viscosity. \(R\) is the so called hydrodynamic ion radius that, in principle, is assumed to be a constant for a given ion-liquid pair.

However, a full-fledged theory for the ion transport in dense gases in a broad density range spanning the crossover region between the gas and liquid state is not yet available. Additionally, there is a lack of experimental data on  negative ions because they are not simply produced by direct ionization of the sample. Actually, the formation of a negative ion is a rather complex process involving low-energy attachment of an excess electron to an electronegative molecular impurity~\cite{Nishikawa1999,Nishikawa1999b,Fehsenfeld1970,Bradbury1933},
%
%~\cite{nishikawa1999b,Nishikawa,Fehsenfeld1970b,Bradbury1933}, %NO~\cite{}, SF\(_6\)~\cite{}, and O\(_2\)~\cite{Bradbury1933},
 thereby leading the formation of a transient anion which may subsequently be stabilized by collision with a third body, typically an atom of the host gas~\cite{christophourou1984a,Matejcik1996,Bloch1935}. 

The stabilization process depends on the ion environment and leads to a ion-medium complex structure that cannot be simply obtained by the adiabatic addition of an ion to the medium. As a result, the ion is located in an empty cavity produced by the short-range repulsive exchange interaction between the excess electron in the ion and the electronic shells of the host gas atoms. In turn, this cavity is surrounded by a solvation shell produced by the electrostriction exerted by the ion on the polarizable medium~\cite{Khrapak1995,Schmidt1999}. The features of the solvation shell depend both on the atomic polarizability of the gas and on its thermodynamic state. Thus, we expect that the structure size is a function of the gas temperature and density.

Over the years, we have conducted extensive experimental investigations on the mobility of oxygen O\(_2^-\) ions in noble gases in broad temperature and density ranges~\cite{Borghesani1993,Borghesani1995,Borghesani1995a,Borghesani1997,Borghesani1999}. We have done several attempts at rationalizing the experimental outcome with quite poor success. For instance, we have either carried out Molecular Dynamics simulations~\cite{Borghesani2008,borghesani2018}, or we extended the hydrodynamic Stokes solution by  accounting for the spatial gradients of density and viscosity induced by electrostriction around the ion~\cite{Borghesani1993}. The most satisfactory results, however, have been obtained by taking into account the variation of the hydrodynamic radius due to the thermodynamic properties of the ion structure and radius~\cite{Khrapak1995,Volykhin1995,Borghesani1997}.

The recent development of a thermodynamic model aimed at predicting the electron bubble radius in superfluid and normal liquid helium and in dense cold helium gas~\cite{Aitken2011,Aitken2011a,Aitken2015,Aitken2016,Aitken2017} in conjunction with the use of the Millikan-Cunningham interpolation formula to bridge the hydrodynamic regime behavior to the dilute gas behavior has allowed us to have a very nice agreement of the theoretical prediction with the experimental data of the O\(_2^-\) ion mobility in dense neon gas even at temperatures close to the critical one~\cite{Borghesani2019,Borghesani2020}, though paying the price of introducing some adjustable parameters.  We want, however, to stress the fact that, if the physical picture underlying the mathematical formulation of the model were uncorrect, the adjustable parameters of the theory, to which we attach physical significance, would not show the coherent and, in some way, predictable behavior we observe.

In view of the success obtained for the case of the neon gas, we exploit the thermodynamic model, also known as the free volume model (henceforth, termed as FVM), to the case of dense argon gas. In this communication we report the experimental data for the O\(_2^-\) ion mobility in argon gas for temperatures \(T\ge 180\,\)K and for densities \(1\times 10^{26}\,\mbox{m}^{-3}\le N\le 50\times 10^{26}\,\mbox{m}^{-3}\) together with the prediction of FVM.

\section{Experimental Results and Discussion}
The experimental apparatus is extensively described in literature~\cite{Borghesani1993} and we briefly recall only the main features. The massive brass cell containing the drift space can withstand pressure up to \(\approx 10\,\)MPa and can be cooled down to \(T=25\,\)K and is thermally stabilized within \(\Delta T=\pm 0.01\,\)K. Pressure is read by a suitable gauge with an accuracy of \(\Delta P \approx \pm 2\,\)kPa. The gas density is computed from \(T\) and \(P\) by means of an accurate equation of state~\cite{wagner}. The drift capacitor is powered by a home-made d.c. generator capable to deliver up to \(3\,\)kV and the drift distance
 is \(\approx 1\,\)cm.

\begin{figure}[h!]
 \centering\includegraphics[width=\columnwidth]{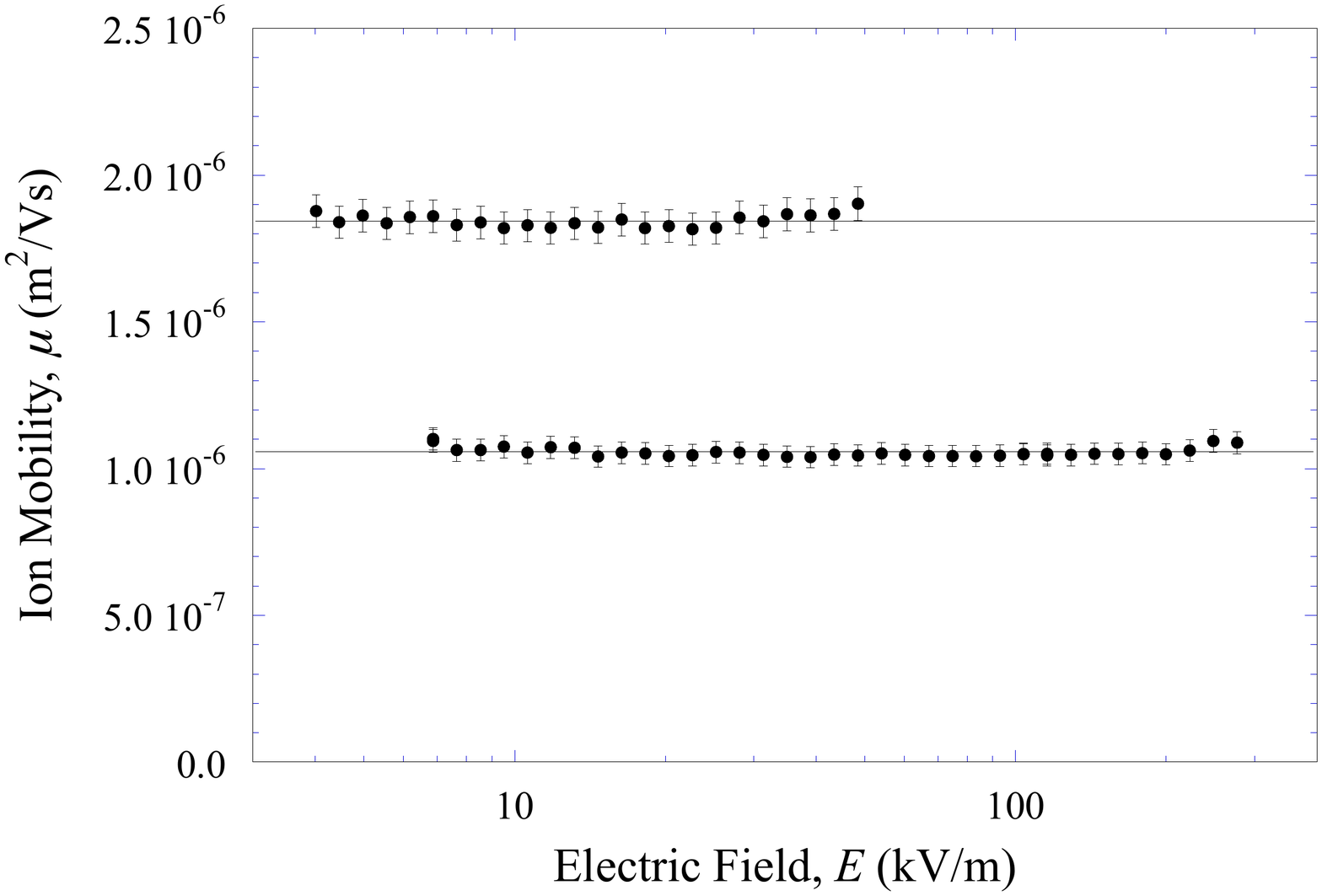}\caption{\small O\(_2^-\) ion mobility \(\mu\) vs electric field \(E\) in argon gas at \(T=260\,\)K for \(N=13.85\times 10^{26}\,\)m\(^{-3}\) (top) and \(N=26.0\times 10^{26}\,\)m\(^{-3}\) (bottom). The solid lines are the respective average values.\label{fig:muvsET260K}}
\end{figure}

Typically, the lowest investigated density is 4 times larger than the ideal gas density at STP (\(N_\mathrm{ig}\approx 2.4\times 10^{25}\,\)m\(^{-3}\) at \(T=273\,\)K and \(P=0.1\,\)MPa).   In this situation, 
 for drift electric fields \(E\) up to \(\approx 300\,\)kV/m,  \(E\) is weak enough not to significantly alter the thermal equilibrium distribution of the ions that are in thermal equlibrium with the gas. As a consequence, the mobility \(\mu\) of thermal ions does not depend on \(E\), as shown in Figure~\ref{fig:muvsET260K}, for all investigated \(N\) and \(T\). 

According to the classical kinetic theory, Eq.\eqref{eq:1}, the density- normalized mobility \(\mu N\) for hard-sphere interaction should be density-independent and should depend on \(T^{-1/2}\). Whereas the former behavior is reasonably well obeyed by the ion mobility in helium gas at \(T=77\,\)K up to intermediate densities~\cite{Khrapak1995}, we have observed significant deviations from this behavior in neon gas~\cite{Borghesani1993,Borghesani1995} and in Argon gas near the critical temperature~\cite{Borghesani1997,Borghesani1999}. We believe that the discrepancy is due to the high gas densities and to the strong electrostriction that the ions exert on the polarizable fluid leading to the formation of a density and temperature dependent ion-medium structure.

\begin{figure}[ht!]
    \centering \includegraphics[width=\columnwidth]{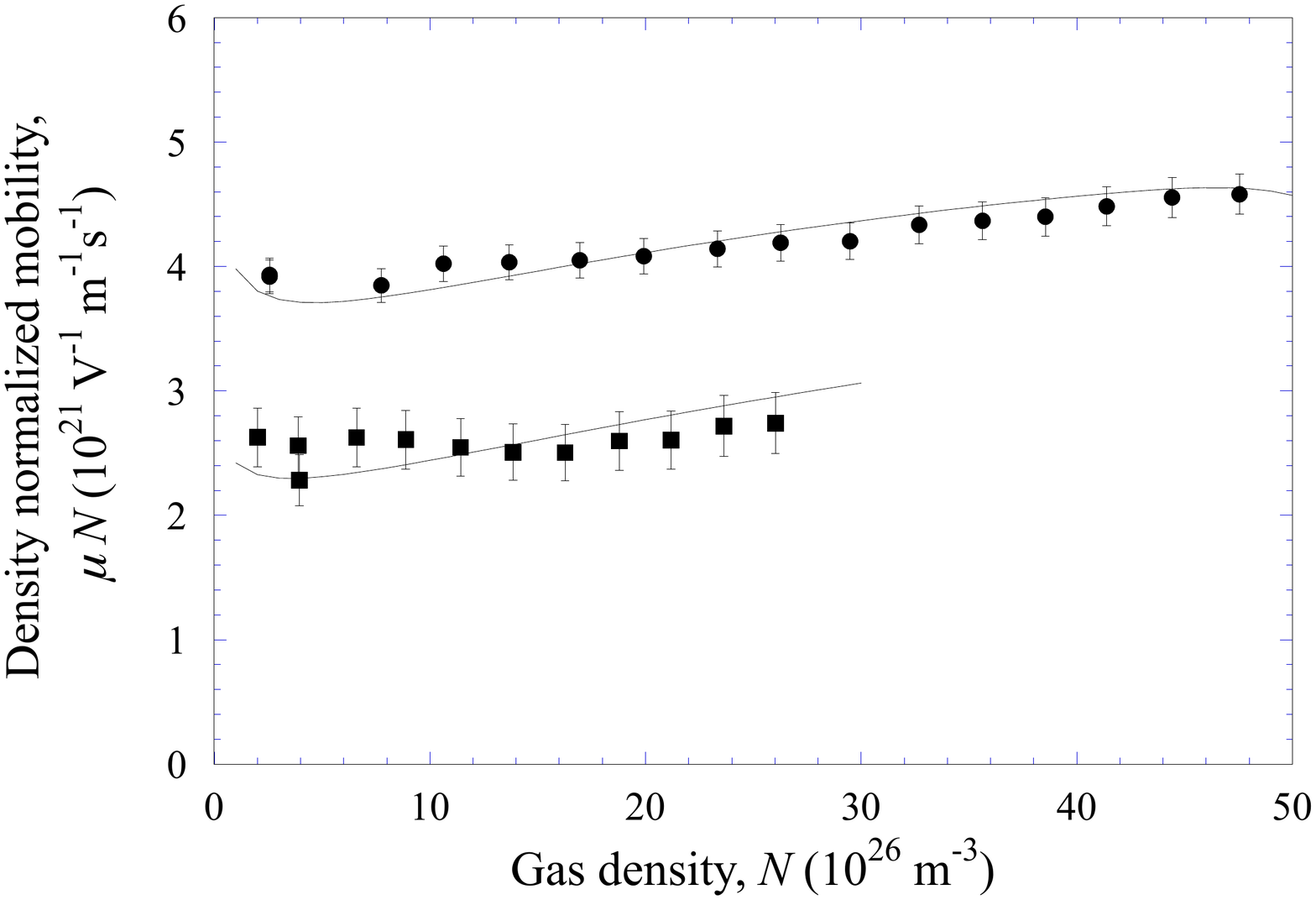}\caption{\small Density normalized mobility \(\mu N\) vs gas density \(N\) for \(T=180\,\)K (upper data) and \(T=260\,\)K (lower data). The solid lines are the prediction of the free volume model.\label{fig:muNvsVT180e260K}}
\end{figure}

In the present experiment we once more note that the density-normalized mobility \(\mu N\) does not follow the prediction of the classical kinetic theory, as can be seen in  Fig.~\ref{fig:muNvsVT180e260K}.
In this figure we report \(\mu N\) as a function of density \(N\) for \(T=180\,\)K and \(T=260\,\)K. \(\mu N\) is neither constant nor shows the predicted \(T^{-1/2}\) behavior. 

At the same time,
 if the pure hydrodynamic approach were valid, Eq.~\ref{eq:2} would  practically yield a straight line passing through zero with quite a steep slope and with almost negligible temperature dependence, in contrast with the experimental results. 

For these reasons, we use the FVM model to rationalize the present experimental results. We anticipate that the model prediction are shown as solid lines in Fig.~\ref{fig:muNvsVT180e260K}.
The FVM, successfully been applied to the case of dense neon~\cite{Borghesani2019,Borghesani2020}, owes its effectiveness to  two reasons: i) 
 it allows the reaserchers to thermodynamically compute the size of the ion-medium structure and ii) it uses the empirical Millikan-Cunningham interpolation formula for the mobility to describe the crossover region between the dense gas- and the hydrodynamic regime.

The FVM yields a thermodynamic description of the free volume \(V_s\) available for the ion motion
\begin{equation}
V_s(T,N)= %\frac
{k_\mathrm{B}T}/\left({P+\Pi}\right)
\label{eq:3}
\end{equation}
\(P\) is the ordinary pressure and \(\Pi\) is the internal pressure that accounts for the attractive potential energy contributions in the system. We expect that the effective ion radius \(R\) is a function of \(V_s\), thereby taking into account also the gas compressibility.

Moreover, the FVM model uses the crossover Millikan-Cunningham formula to describe the mobility in the dense gas region
\begin{equation}
\mu N= \frac{eN}{6\pi \eta R }\left\{1+\varphi\left[K_n(N,T)\right]
\right\}\label{eq:4}\end{equation}\(\varphi\) is the slip correction factor that is a function of the Knudsen number \(K_n=\ell/R\), where \(\ell=\) is the ion mean free
 path
 \begin{equation}
 \ell (N,T) = \frac{3\eta}{N \left(8m_r k_\mathrm{B}T/\pi\right)^{1/2} }
 \label{eq:4b}
 \end{equation}
 
 We found that the internal pressure can be cast in the form \(\Pi=\alpha N^2\) with \(\alpha= 8.937\times 10^5\,\)MPa nm\(^6\) is an universal constant valid for all noble gases~\cite{Aitken2011,Borghesani2019}.  We also found that the hydrodynamic radius has the general expression
 \begin{equation}
 \frac{R}{R_0} =1 + \frac{\left(V_0/V_s\right)^{\epsilon_1}}{1+\gamma \left(V_0/V_s\right) }
 \label{eq:5}
 \end{equation} with \(\gamma=2\) and \(\epsilon_1= 12\). \(R_0=0.55\,\)nm is the hard-sphere radius of the charge-medium interaction and depends on the medium nature. \(V_0=\delta k_\mathrm{B}T\)  is the  free volume of a suitable reference state. The value \(\delta =0.15\) appears to be universal among noble gases.

 Finally, the slip correction factor can be cast in the form
 \begin{equation}
 \varphi(N,T)= \frac{2f}{3}\left(\frac{N_c}{N}\right)^{1.1}\exp{\left[-0.5/{K_n}^{1.8}+T_c/T
 \right]}
 \label{eq:6}
 \end{equation}
 in which \(N_c\) and \(T_c\) are the critical temperature and density of the gas, and \(f\approx 1\) is an adjustable parameter.

The adjustable parameters in the model are determined by fitting the data at a given temperature, in our case \(T=180\,\)K, and keep the same  values on all other temperatures. 
The FVM predictions for some isotherms are shown as solid lines in Fig.~\ref{fig:muNvsVT180e260K} and are in reasonably good agreement with the experimental data. Similar results have been obtained for all other investigated temperatures.

\begin{figure}[ht!]\centering\includegraphics[width=\columnwidth]{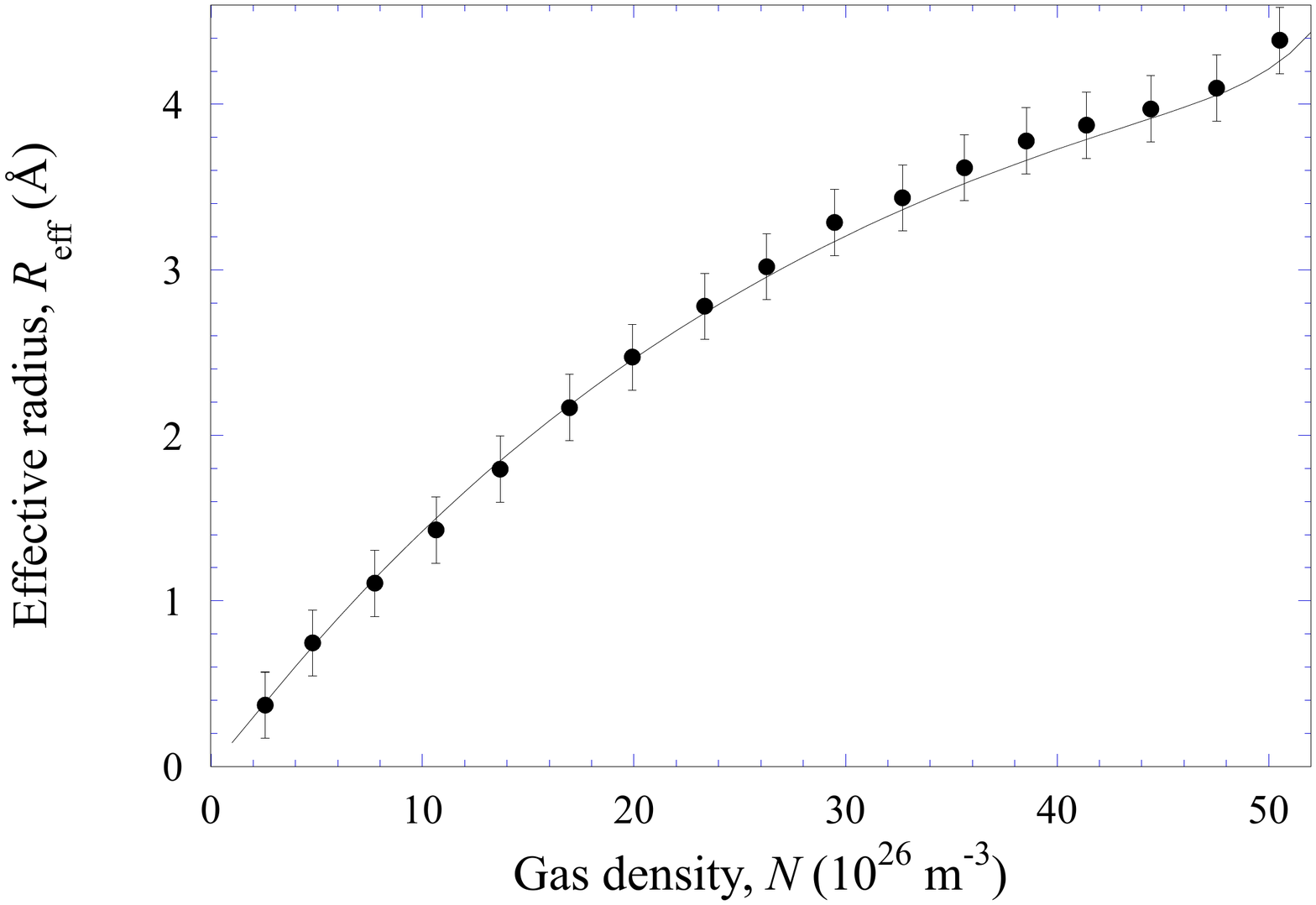}\caption{\small Effective hydrodynamic radius \(R_\mathrm{eff}\) vs gas density \(N\)  for \(T=180\,\)K. Points: experimental data. Line: theoretical prediction.\label{fig:reff180}}
\end{figure}

The effective hydrodynamic radius can be obtained as
\begin{equation}
R_\mathrm{eff} = \frac{R}{1+\varphi(N,T)}
\label{eq:7}\end{equation}
and is plotted for \(T=180\,\)K in Fig.~\ref{fig:reff180} because it is the isotherm for which the density range is the broadest. The model prediction (solid line) is in very good agreement with the effective radius determined by inverting the experimental mobility data. At all other temperatures we obtain similar results and the small differences we observed between the different isotherms are mainly due to the temperature dependence of the gas viscosity.
We note that the effective radius depends almost linearly on the gas density in the low density region. This implies that \(\mu N\) at low density is roughly independent of \(N\), as can be expected from classical kinetic theory. However, as soon as the density is increased above, say, \(N\simeq 10\times 10^{26}\,\)m\(^{-3}\), the effective radius significantly deviates from linearity and from the expectations of a binary collision limited mobility. We believe that the crossover region towards the hydrodynamic behavior of the mobility is shifted to quite low densities in argon because of its large polarizability. 
The strong electrostriction effect leads to a larger and more density dependent ionic-medium complex and the perturbation  induced by the ionic charge extends further away in the gas, thereby making the binary-collision picture of the classical kinetic theory fail earlier than expected.

\begin{figure}[ht!]
     \centering\includegraphics[width=\columnwidth]{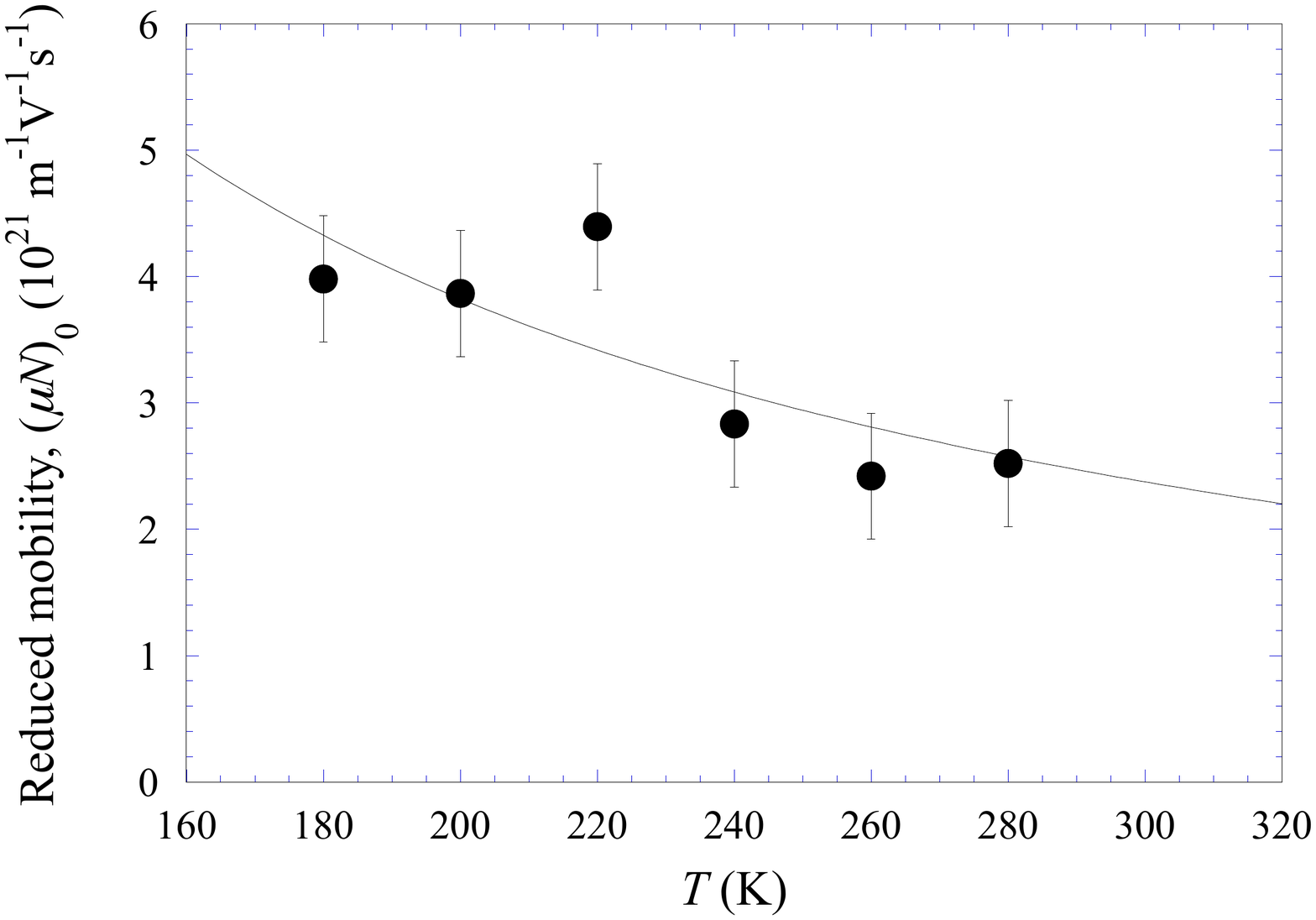}\caption{\small Reduced mobility \(\mu_\mathrm{red}=\left(\mu N\right)_0\)vs \(T\). Points: Experimental data. Solid line: \(1/T\) fit to the data.\label{fig:mu0novsT}}
 \end{figure}

 In Fig.~\ref{fig:mu0novsT} we plot the so-called reduced mobility, i.e., the zero-density limit of the density-normalized mobility \(\mu_\mathrm{red} \equiv\left(\mu N\right)_0 = \lim_{N\rightarrow 0} \mu N\). The experimental data are quite scattered but are roughly fitted to a \(T^{-1}\) curve. Once more this observed behavior disagrees with the prediction of classical kinetic theory. 
\((\mu N)_0\) should be \(\propto T^{-1/2}\) in case of binary collision of hard-spheres and should be independent of \(T\) if the low-energy ion-atom interaction were dominated by the long-range polarization interaction. The observed experimental behavior suggests that even at the lowest investigated density the condition of dilute gas is never satisfied: in such a highly polarizable gas the electrostriction phenomenon inhibits the observation of the bare ion-atom binary collisions. 

\begin{figure}[ht!]
     \centering
     \includegraphics[width=\columnwidth]{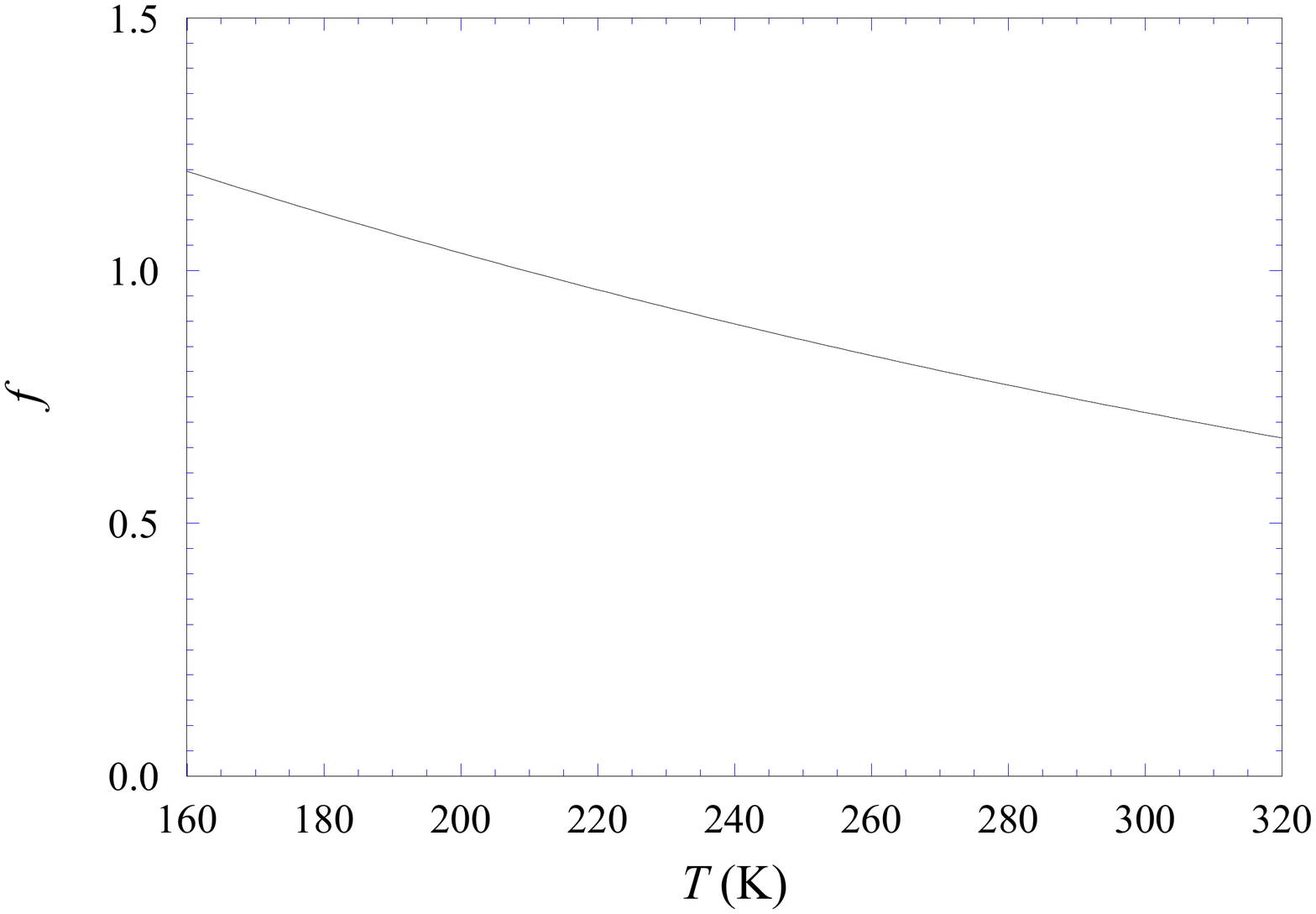}\caption{\small Temperature dependence of the adjustable factor \(f\) of the slip correction factor. \label{fig:fFactorvsT}}
 \end{figure}

Finally, the FVM is able to reproduce the fitted \(T^{-1}\) dependence of the reduced mobility if the ajustable factor \(f\) in Eq.~\ref{eq:7} depends on \(T\) in the way shown in Fig.~\ref{fig:fFactorvsT}.

\section{Conclusions}
We have reported measurements of the O\(_2^-\) ion mobility in argon gas up to moderate densities for several temperatures from \(T=180\,\)K up to room temperature in a density range that spans the crossover region between the dilute gas regime and the hydrodynamic one. We have shown that neither the classical kinetic theory nor the pure hydrodynamic theory are able to satisfactorily rationalize the data. 

By contrast, we have shown that the Millikan-Cunningham slip correction to the Stokes hydrodynamic formula for the mobility give a reasonable agreement with the data if the effective ion radius is computed by means of the free volume model that is a thermodynamic model aimed at including in a van-der-Waals-like approach the attractive potential energy contributions in the system due to the long-range ion-atom interaction. 

The FVM analysis of the O\(_2^-\) ion mobility  data close to the critical temperature in argon is at present underway.
 %\vglue 1em
%\section*{Acknowledgment}
%\bibliographystyle{unsrt}
\bibliographystyle{IEEEtran} 
%\bibliography{FVM.bib}

\begin{thebibliography}{10}

\bibitem{Bruggeman2016b}
P~J Bruggeman and {\em et al.}
\newblock {Plasma-liquid interactions: A review and roadmap}.
\newblock {\em Plasma Sources Sci. Technol.}, 25:053002, 2016.

\bibitem{lopez2005}
I.~M. Lopez and V.~Chepel.
\newblock {\em {Electronic Excitations in Liquefied Rare Gases}}, chapter {Rare
  Gas Liquid Detectors}, pages 331--388.
\newblock American Scientific Publishers, Stevenson Ranch, CA (USA), 2005.

\bibitem{mason2001}
P.~Hughes and N.~Mason.
\newblock {\em {Introduction to Environmental Physics: Planet Earth, Life and
  Climate}}.
\newblock CRC Press, Boca Raton, 2001.

\bibitem{viehland1975}
L.~A. Viehland and E.~A. Mason.
\newblock {Tables of transport collision integrals for \((n,6,4\) ion-neutral
  potentials}.
\newblock {\em At. Data and Nucl. Data Tables}, 16:495--514, 1975.

\bibitem{viehland1975b}
L.~A. Viehland and E.~A. Mason.
\newblock {Gaseous ion mobility in electric fields of arbitrary strength}.
\newblock {\em Ann. Phys.}, 91:499--533, 1975.

\bibitem{maitland}
G.~C. Maitland, M.~Rigby, and W.~A. Wakeham.
\newblock {\em {Intermolecular Forces. Their Origin adn Determination}}.
\newblock Clarendo Press, Oxford, 1981.

\bibitem{mason1988}
E.~A. Mason and E.~W. McDaniel.
\newblock {\em {Transport Properties of Ions in Gases}}.
\newblock Wiley, New York, 1988.

\bibitem{Nishikawa1999}
M.~Nishikawa, R.A. Holroyd, and K.~Itoh.
\newblock {Electron Attachment to NO in Supercritical Ethane}.
\newblock {\em J. Phys. Chem. B}, 102:4189--4192, 1998.

\bibitem{Nishikawa1999b}
M.~Nishikawa, K.~Itoh, and Richard~A. Holroyd.
\newblock {Electron Attachment to CO\(_2\) in Supercritical Ethane}.
\newblock {\em J. Phys. Chem. A}, 103:550--556, 1999.

\bibitem{Fehsenfeld1970}
F~C Fehsenfeld.
\newblock {Electron Attachment to SF\(_6\)}.
\newblock {\em J. Chem. Phys.}, 53:2000--2004, 1970.

\bibitem{Bradbury1933}
N.~E. Bradbury.
\newblock {Electron attachment and negative ion formation in oxygen and oxygen
  mixtures}.
\newblock {\em Phys. Rev.}, 44:883--890, 1933.

\bibitem{christophourou1984a}
L.~G. Christophorou, D.~L. McCorkle, and A.~A. Christodoulides.
\newblock {\em {Electron-Molecule Interactions and Their Applications}},
  volume~I, chapter {Electron Attachment Processes}.
\newblock Academic Press, Orlando, 1984.

\bibitem{Matejcik1996}
S.~Matejcik, A.~Kiendler, P.~Stampfli, A.~Stamatovic, and T.~D. M{\"{a}}rk.
\newblock {Vibrationally Resolved Electron Attachment to Oxygen Clusters}.
\newblock {\em Phys. Rev. Lett.}, 77:3771--3774, 1996.

\bibitem{Bloch1935}
F.~Bloch and N.~E. Bradbury.
\newblock {On the mechanism of unimolecular electron capture}.
\newblock {\em Phys. Rev.}, 48:689--695, 1935.

\bibitem{Khrapak1995}
A~G Khrapak, W~F Schmidt, and K~F Volykhin.
\newblock {Structure of O2- in dense helium gas}.
\newblock {\em Phys. Rev. E}, 51:4804--4806, 1995.

\bibitem{Schmidt1999}
W.~F. Schmidt, K.~F. Volykhin, A.~G. Khrapak, and E.~Illenberger.
\newblock {Structure and mobility of positive and negative ions in non-polar
  liquids}.
\newblock {\em J. Electrostat.}, 47:83--95, 1999.

\bibitem{Borghesani1993}
A.~F. Borghesani, D.~Neri, and M~Santini.
\newblock {Low-temperaure O2- mobility in high-density neon gas}.
\newblock {\em Phys. Rev. E}, 48:1379--1389, 1993.

\bibitem{Borghesani1995}
A.~F. Borghesani, F.~Chiminello, D.~Neri, and M.~Santini.
\newblock {O- 2 ion mobility in compressed He and Ne Gas}.
\newblock {\em Int. J. Thermophys.}, 16:1235--1244, 1995.

\bibitem{Borghesani1995a}
A~F Borghesani, F.~Chiminello, D~Neri, and M~Santini.
\newblock {O- 2 ion mobility in compressed He and Ne Gas}.
\newblock {\em Int. J. Thermophys.}, 16:1235--1244, 1995.

\bibitem{Borghesani1997}
A.~F. Borghesani, D.~Neri, and A.~Barbarotto.
\newblock {Mobility of 02- ions in near critical Ar gas}.
\newblock {\em Chem. Phys. Lett.}, 267:116--122, 1997.

\bibitem{Borghesani1999}
A.F. Borghesani, D~Neri, and A.~Barbarotto.
\newblock {Critical behavior of O2 - ions in Argon gas}.
\newblock {\em Int. J. Thermophys.}, 20:899--909, 1999.

\bibitem{Borghesani2008}
A.~F. Borghesani.
\newblock {Mobility of O2 - ions in supercritical Ar: Experiment and molecular
  dynamics simulations}.
\newblock {\em Int. J. Mass Spectrom.}, 277:220--222, 2008.

\bibitem{borghesani2018}
A.~F. Borghesani and F.~Aitken.
\newblock {Molecular dynamics simulations of the O\(_2^-\) ion mobility in
  dense Ne gas at low temperature: Influence of the repulsive part of the
  ion-neutral interaction potential}.
\newblock {\em IEEE Trans. Dielectr. Electr. Insul.}, 25:1992--1998, 2018.

\bibitem{Volykhin1995}
K~F Volykhin and A~G Khrapak.
\newblock {Structure and mobility of negative ions in dense gases and nonpolar
  liquids}.
\newblock {\em JETP}, 81:901--908, 1995.

\bibitem{Aitken2011}
F.~Aitken, Z-L Li, N.~Bonifaci, A.~Denat, and K.~von Haeften.
\newblock {Electron mobility in liquid and supercritical helium measured using
  corona discharges: a new semi-empirical model for cavity formation.}
\newblock {\em Phys. Chem. Chem. Phys.}, 13:719--724, 2011.

\bibitem{Aitken2011a}
F.~Aitken, N.~Bonifaci, A.~Denat, and K.~{Von Haeften}.
\newblock {A macroscopic approach to determine electron mobilities in
  low-density helium}.
\newblock {\em J. Low Temp. Phys.}, 162:702--709, 2011.

\bibitem{Aitken2015}
Fr. Aitken, N.~Bonifaci, L.~G. Mendoza-Luna, and K.~von Haeften.
\newblock {Modelling the mobility of positive ion clusters in normal liquid
  helium over large pressure ranges}.
\newblock {\em Phys. Chem. Chem. Phys.}, 17:18535--18540, 2015.

\bibitem{Aitken2016}
Fr. Aitken, N.~Bonifaci, K.~{Von Haeften}, and J.~Eloranta.
\newblock {Theoretical modeling of electron mobility in superfluid 4He}.
\newblock {\em J. Chem. Phys.}, 145(4):044105, 2016.

\bibitem{Aitken2017}
F.~Aitken, F.~Volino, L.G. Mendoza-Luna, K.V. Haeften, and J.~Eloranta.
\newblock {A thermodynamic model to predict electron mobility in superfluid
  helium}.
\newblock {\em Phys. Chem. Chem. Phys.}, 19:15821--15832, 2017.

\bibitem{Borghesani2019}
A~F Borghesani and F~Aitken.
\newblock A thermodynamic model for o\(_2^-\) mobility in neon gas over broad
  density and temperature ranges.
\newblock {\em Plasma Sources Science and Technology}, 28:03LT01, 2019.

\bibitem{Borghesani2020}
A.~F. Borghesani and F.~Aitken.
\newblock O\(_2^-\) ion mobility in dense ne gas: The free volume model.
\newblock {\em IEEE Trans. Dielectr. Electr. Insul.}, 27:757--763, 2020.

\bibitem{wagner}
Ch. Tegeler, R.~Span, and W.~Wagner.
\newblock {A New Equation of State for Argon Covering the Fluid Region for
  Temperatures From the Melting Line to 700 K at Pressures up to 1000 MPa}.
\newblock {\em J. Phys. Chem. Ref. Data}, 28:779--850, 1999.

\end{thebibliography}

\end{document}